\documentclass[twocolumn]{jpsj2} 
\title{Abrupt Emergence of Pressure-Induced Superconductivity of 34 K in SrFe$_2$As$_2$:\\ A Resistivity Study under Pressure}

\author{Hisashi \textsc{Kotegawa}$^{1,2}$\thanks{E-mail address: kotegawa@crystal.kobe-u.ac.jp}, Hitoshi \textsc{Sugawara}$^{3}$, and Hideki \textsc{Tou}$^{1,2}$}

\inst{$^{1}$Department of Physics, Kobe University, Kobe 658-8530\\
$^{2}$JST, Transformative Research-Project on Iron Pnictides (TRIP), Chiyoda, Tokyo 102-0075\\
$^{3}$Faculty of Integrated Arts and Science, Tokushima University, Tokushima 770-8502\\
}

\abst{
We report resistivity measurement under pressure in single crystals of SrFe$_2$As$_2$, which is one of the parent materials of Fe-based superconductors. The structural and antiferromagnetic (AFM) transition of $T_0 = 198$ K at ambient pressure is suppressed under pressure, and the ordered phase disappears above $P_c \sim 3.6-3.7$ GPa. Superconductivity with a sharp transition appears accompanied by the suppression of the AFM state. $T_c$ exhibits a maximum of 34.1 K, which is realized close to the phase boundary at $P_c$. This $T_c$ is the highest among those of the stoichiometric Fe-based superconductors.
}

\kword{SrFe$_2$As$_2$, superconductivity, pressure, single crystal}

\begin{document}
\maketitle

After the discovery of superconductivity at 26 K in F-doped system LaFeAsO$_{1-x}$F$_x$ (ZrCuSiAs-type structure),\cite{Kamihara} various Fe-based materials have been reported to show superconductivity.\cite{Rotter,Sasmal,Wang,Hsu}
Among them, $A$Fe$_2$As$_2$ ($A=$ Ca, Sr, and Ba) systems with a ThCr$_2$Si$_2$-type structure show superconductivity by doping with K or Cs into the $A$ site,\cite{Rotter,Sasmal} or by doping Co into the Fe site.\cite{Sefat,Leithe-Jasper}
Doping is an effective method of inducing superconductivity in Fe-based superconductors.
However, this simultaneously induces the inhomogeneity of the crystal structure and electronic state.
The inhomogeneity sometimes makes it difficult to observe the intrinsic properties of the material.

Instead of doping, the application of pressure for undoped compound is also an effective method of inducing superconductivity.
Pressure-induced superconductivity in $A$Fe$_2$As$_2$ ($A=$ Ca, Sr, and Ba) has been reported.\cite{Torikachvili,Park,Alireza}
The superconductivity of these stoichiometric compounds is important for the study of Fe-based superconductors.
Concerning CaFe$_2$As$_2$, its superconductivity has been recognized to be intrinsic, because some groups have reported that the zero-resistance state is observed in a similar pressure range.\cite{Torikachvili,Park,Lee}
In the cases of BaFe$_2$As$_2$ and SrFe$_2$As$_2$, Alireza {\it et al.} have reported that Meissner effects appear between $2.5-6.0$ GPa for BaFe$_2$As$_2$ and between $2.8-3.6$ GPa for SrFe$_2$As$_2$ using magnetization measurements under pressure.\cite{Alireza}
However, Fukazawa {\it et al.} have observed no zero-resistance state at pressures of up to 13 GPa in BaFe$_2$As$_2$.\cite{Fukazawa}
On the other hand, Kumar {\it et al.} performed resistivity measurement at pressures of up to 3 GPa in SrFe$_2$As$_2$ and reported that the onset of superconductivity appears above 2.5 GPa, but they observed no zero-resistance state up to 3 GPa.\cite{Kumar}
Quite recently, Igawa {\it et al.} have reported that the zero-resistance state was realized below 10 K at a high pressure of 8 GPa in SrFe$_2$As$_2$, but that the transition was broad.\cite{Igawa}
No consensus on pressure-induced superconductivity in BaFe$_2$As$_2$ and SrFe$_2$As$_2$ has been arrived at yet.

In this paper, we report the results of resistivity measurements in single-crystalline samples of SrFe$_2$As$_2$ up to 4.3 GPa.
This is the first resistivity measurement above 3 GPa using single-crystalline samples.
In our measurements, the zero-resistance state below $T_c=34$ K with a sharp transition was observed above 3.5 GPa.

\begin{figure}[htb]
\centering
\includegraphics[width=0.9\linewidth]{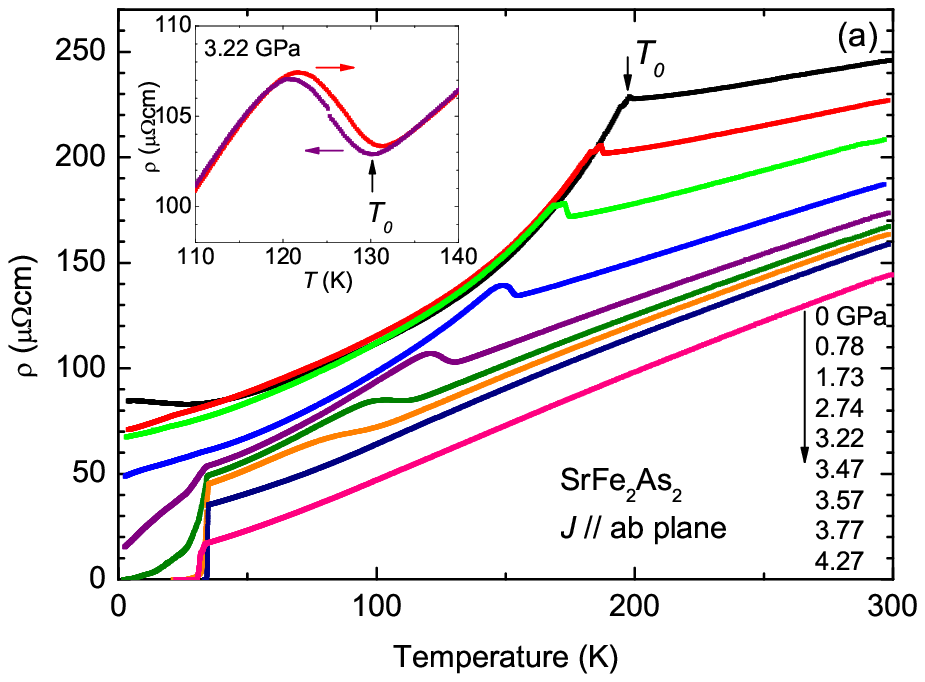}
\includegraphics[width=0.9\linewidth]{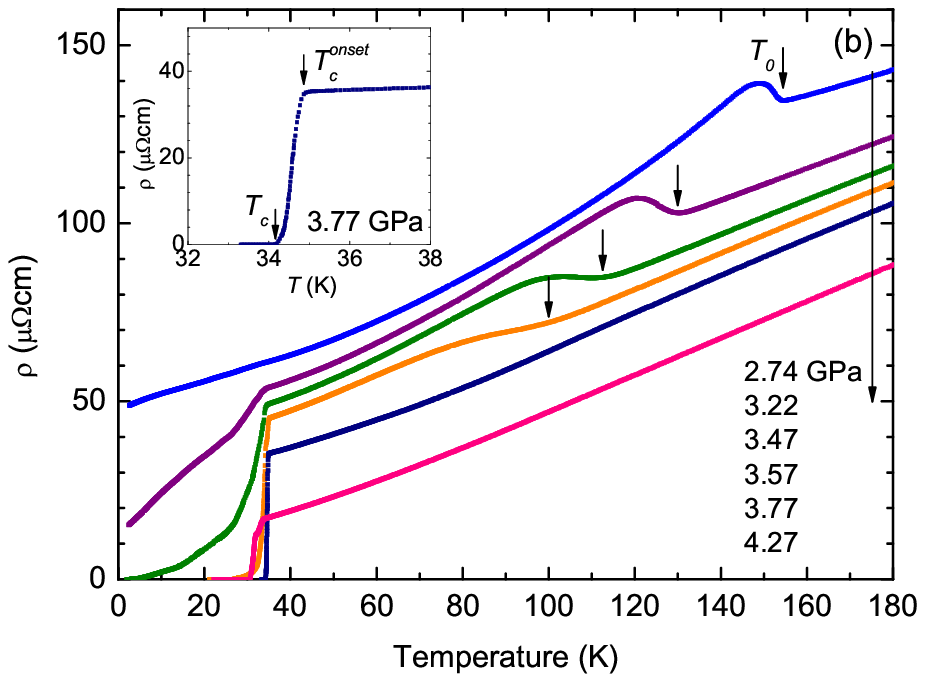}
\caption[]{
(color online) Temperature dependence of the in-plane resistivity in SrFe$_2$As$_2$ below (a) 300 and (b) 180 K. The arrows indicate the structural and AFM phase transition temperature $T_0$. The transition disappears above $P_c\sim 3.6-3.7$ GPa. The inset of Fig.~(a) shows the hysteresis around $T_0$ for the temperature history at 3.22 GPa. Superconductivity with zero resistance is observed above approximately $P_c$. The maximum $T_c$ is 34.1 K at 3.77 GPa, as shown in the inset of Fig.~(b).
}
\end{figure}

Single-crystalline samples were prepared by the Sn-flux method as reported in ref. 15.
Electrical resistivity ($\rho$) measurement at high pressures was carried out using an indenter cell.\cite{indenter}
$\rho$ was measured by a four-probe method while introducing a flow of current along the $ab$ plane.
Daphne oil 7373 was used as a pressure-transmitting medium.
Applied pressure was estimated from the $T_{c}$ of the lead manometer.
Resistivity measurement under pressure was performed for two settings using different samples and almost the same results were obtained between two samples.

Figures 1(a) and 1(b) show the temperature dependences of $\rho$ at several pressures of up to 4.3 GPa.
A clear anomaly was observed at 198 K at ambient pressure, which is similar to that of Yan {\it et al.}'s sample.\cite{Yan} 
This temperature, denoted as $T_0$, corresponds to the structural transition temperature and the simultaneous magnetic transition temperature.\cite{Krellner,Yan,Zhao}
The magnetic structure of SrFe$_2$As$_2$ has been reported to be a collinear antiferromagnetic (AFM) one.\cite{Zhao}
The $T_0$ of our sample is lower than that of Kumar {\it et al.}'s sample.\cite{Kumar}
$\rho$ shows a small jump at $T_0$ in our sample and the jump becomes remarkable under pressure, in contrast to other measurements under pressure.\cite{Kumar,Igawa}
The reason why the jump appears in our sample is unclear at present, but this behavior is understood to be induced by the reconstruction of the Fermi surface owing to the AFM transition, and resembles that of CaFe$_2$As$_2$.\cite{Torikachvili,Park}
Thus, we define the temperature at the jump as $T_0$ on the analogy of CaFe$_2$As$_2$, as shown in Fig.~1(b).
As shown in the figure, $T_0$ decreases with increasing pressure and reaches $\sim100$ K at 3.57 GPa.
No signature of the transition at $T_0$ was observed above 3.77 GPa, indicating the disappearance of the AFM state.
The critical pressure between the AFM state and the paramagnetic (PM) state is estimated to be $P_c\sim 3.6-3.7$ GPa.
The inset of Fig.~1(a) displays $\rho(T)$ of around $T_0$ at 3.22 GPa.
A small hysteresis was observed between cooling and warming, indicative of the first-order phase transition.

The onset of superconductivity appears above $\sim3$ GPa but the transition is quite broad, similarly to that observed in the experiments by Kumar {\it et al.}.\cite{Kumar}
A zero-resistance state is observed above 3.47 GPa, and the transition becomes sharper above 3.77 GPa where the AFM state is no longer realized.
In this paper, $T_c$ is defined by the temperature of the zero resistance.
The maximum $T_c$ was 34.1 K at 3.77 GPa, as shown in the inset of Fig.~1(b).
This $T_c$ is close to $37-38$ K of the doped systems (Ba$_{0.6}$K$_{0.4}$)Fe$_2$As$_2$ and (K$_{0.4}$Sr$_{0.6}$)Fe$_2$As$_2$.\cite{Rotter,Sasmal} 
Above 3.77 GPa, $T_c$ is almost constant but slightly decreases with increasing pressure.

\begin{figure}[htb]
\centering
\includegraphics[width=0.9\linewidth]{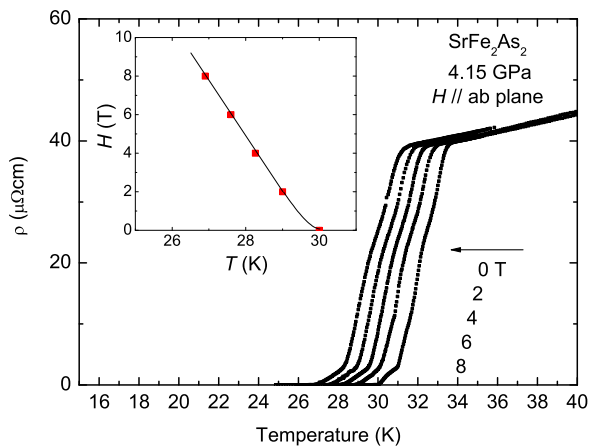}
\caption[]{
(color online) Temperature dependence of resistivity under magnetic fields of 0, 2, 4, 6, and 8 T in SrFe$_2$As$_2$ under 4.15 GPa. The inset shows the field dependence of $T_c$. The initial slope is estimated to be $-0.35$ K/T.
}
\end{figure}

Figure 2 shows $\rho(T)$ under magnetic field at 4.15 GPa, when the magnetic field was applied along the $ab$-plane.
$T_c$ decreases from 30 K at 0 T to $\sim27$ K at 8 T.
The initial slope was estimated to be $-0.35$ K/T, giving $H_{c2}\sim 86$ T by linear extrapolation.
These values are comparable to those of other Fe-based compounds.

\begin{figure}[b]
\centering
\includegraphics[width=0.95\linewidth]{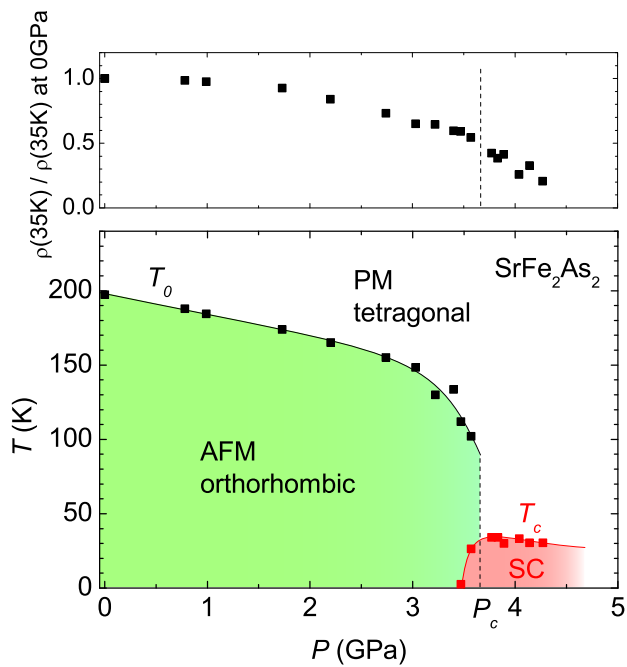}
\caption[]{
(color online) Pressure-temperature phase diagram for SrFe$_2$As$_2$ and pressure-dependence of $\rho$ at 35 K. $T_0$ decreases with application of pressure, and the slope becomes steeper above $\sim3$ GPa. The magnetically ordered phase most likely disappears at around $P_c \sim 3.6-3.7$ GPa. There is no distinct anomaly in $\rho(35{\rm K})$ at around $P_c$. Superconductivity appears above 3.5 GPa accompanied by the suppression of the AFM state. Information on the crystal structure was obtained from ref. 13.
}
\end{figure}

Figure 3 shows the pressure-temperature phase diagram of SrFe$_2$As$_2$.
The initial slope of $T_0$ was estimated to be $dT_0/dP \sim -13$ K/GPa, which is the same as that of Kumar {\it et al.}.\cite{Kumar}
The ordered phase was markedly suppressed above 3 GPa, and no signature of the AFM state was observed at 3.77 GPa.
The ordered state up to 3 GPa is confirmed to have an orthorhombic crystal structure.\cite{Kumar}
The superconductivity appears from slightly below $P_c \sim 3.6-3.7$ GPa, and exhibits the highest $T_c = 34.1$ K in the PM state close to $P_c$.

In CaFe$_2$As$_2$, another structural phase transition from the tetragonal phase to the "collapsed" tetragonal one has been reported under high pressure,\cite{Kreyssig} which can be detected by $\rho(T)$.\cite{Lee} 
In contrast, there is no corresponding distinct anomaly above $P_c$ in SrFe$_2$As$_2$.
In CaFe$_2$As$_2$, the pressure dependence of the residual resistivity indicates the anomalous behavior of a dome shape.\cite{Lee}
We plot the pressure-dependence of $\rho$ at 35 K for SrFe$_2$As$_2$ in the upper panel of the figure, but $\rho(35{\rm K})$ shows a gradual decrease under pressure, and no anomalous behavior was observed for SrFe$_2$As$_2$.

\begin{figure}[htb]
\centering
\includegraphics[width=0.9\linewidth]{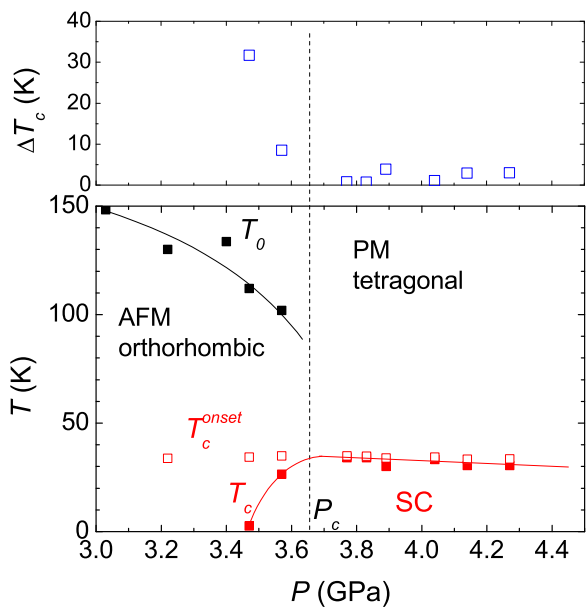}
\caption[]{
(color online) Pressure-temperature phase diagram for SrFe$_2$As$_2$ at around $P_c$ and pressure dependence of $\Delta T_c = T_c^{onset}-T_c$. The superconducting transition is sharp above $P_c$, whereas $\Delta T_c$ is wide below $P_c$. $T_c^{onset}$ is almost independent of pressure.
}
\end{figure}

Figure 4 is the pressure-temperature phase diagram around the phase boundary.
We plotted the onset temperature of superconductivity, $T_c^{onset}$, and the transition width, $\Delta T_c=T_c^{onset}-T_c$.
The zero-resistance state is observed even in the narrow pressure range below $P_c$.
In the Fe-based superconductors, it is a controversial issue whether superconductivity can coexist with the AFM state.\cite{Luetkens,Chen,Drew}
Since the resistivity is macroscopic measurement and is sensitive to superconductivity, it is generally difficult to discuss this issue. 
However, note that $\Delta T_c$ is unusually wide below $P_c$.
In contrast, $\Delta T_c$ becomes markedly sharper above $P_c$.
The minimum $\Delta T_c$ is 0.75 K at 3.83 GPa.
This indicates that the PM state favors superconductivity and that the AFM state prevents the occurrence of superconductivity in SrFe$_2$As$_2$.
The superconductivity with a wide $\Delta T_c$ below $P_c$ implies non bulk superconductivity.
The transition from the tetragonal structure to the orthorhombic one is of the first order.\cite{Krellner}
As seen in the inset of Fig.~1(a), the transition at $T_0$ is of the first order even close to $P_c$.
At high pressures and low temperatures, the pressure distribution is inevitable.
If the transition at $P_c$ is of the first order, the pressure distribution is expected to induce phase separation.
We speculate that the observed superconductivity below $P_c$ originates from the phase-separated PM phase.
This is supported by the fact that $T_c^{onset}$ is almost independent of pressure below $P_c$.
However, if the phase separation is realized at around $P_c$, we expect the enhancement of $\rho$ at low temperatures at around $P_c$ owing to scattering at the domain boundary.
As shown in Fig.~3, there is no anomalous behavior in $\rho(35{\rm K})$ at around $P_c$ within experimental error.
The phase separation and coexistence of superconductivity and magnetism are still an open question, and confirmation by microscopic measurements is required.

To our knowledge, the pressure-temperature phase diagram of SrFe$_2$As$_2$ has been reported by three groups.\cite{Alireza,Kumar,Igawa}
Our phase diagram is almost consistent with that of Kumar {\it et al.}, although their resistivity measurements have been performed only up to 3 GPa.\cite{Kumar}
On the other hand, the phase diagrams by Alireza {\it et al.} and Igawa {\it et al.} are different from ours.
Alireza {\it et al.} have used a single-crystalline sample and Daphne oil 7373 as a pressure transmitting medium, which are the same as those used in our measurements.
In their phase diagram, the superconductivity of $T_c \sim 27$ K appears abruptly at 2.8 GPa and disappears above 3.6 GPa.
The pressure region of superconductivity is quite different.
On the other hand, Igawa {\it et al.} have used a polycrystalline sample, and Fluorinert (FC-77:FC-70 = 1:1) and NaCl as a pressure transmitting medium.
The onset of superconductivity was observed in a wide pressure range, and zero resistance below 10 K was realized at a high pressure of 8 GPa.
The $T_c^{onset}$ at around $3-4$ GPa is almost the same as that in our measurements, but the zero-resistance state is different.
In their phase diagram, the AFM state is drawn to survive up to 8 GPa.
The differences between samples and/or pressure-transmitting mediums are considered to induce the inconsistency between the obtained phase diagrams.

To summarize, we have investigated the resistivity under pressure in a single-crystalline SrFe$_2$As$_2$ up to 4.3 GPa.
According to our resistivity measurement, the magnetically ordered phase most likely disappears abruptly above $P_c \sim 3.6-3.7$ GPa, and superconductivity appears above approximately $P_c$; however, other experimental methods are required to confirm whether this phase diagram reflects bulk properties.
The maximum $T_c$ was 34.1 K for the pressure-induced superconductivity in stoichiometric SrFe$_2$As$_2$, which is close to $37-38$ K of the doped systems.\cite{Rotter,Sasmal}
The maximum $T_c$ is realized in the PM state close to $P_c$.
This gives us two different scenarios.
One is that the instability of the AFM state plays an important role in superconductivity.
Another is that the AFM state obstructs the optimized situation for higher $T_c$.
Systematic investigations are needed to elucidate the relation between superconductivity and magnetism, but the stoichiometric system SrFe$_2$As$_2$ is a good candidate for treating this issue.

We thank Y. Hara and T. Kawazoe for experimental assistance.
This work has been partly supported by Grant-in-Aids for Scientific Research (Nos. 19105006, 19204036, 19014016, and 20045010) from the Ministry of Education, Culture, Sports, Science, and Technology (MEXT) of Japan.

\end{document}